# The Information Ecosystem of Conspiracy Theory: Examining the QAnon Narrative on Facebook

SOOJONG KIM, University of California Davis, Davis, United States

JISU KIM, Singapore Institute of Technology, Singapore, Singapore

There has been concern about the proliferation of the "QAnon" conspiracy theory on Facebook, but little is known about how its misleading narrative propagated on the world's largest social media platform. Thus, the present research analyzed content generated by 2,813 Facebook pages and groups that contributed to promoting the conspiracy narrative between 2017 and 2020. The result demonstrated that activities of QAnon pages and groups started a significant surge months before the 2020 U.S. Presidential Election. We found that these pages and groups increasingly relied on internal sources, i.e., Facebook accounts or their content on the platform, while their dependence on external information sources decreased continuously since 2017. It was also found that QAnon posts based on the Facebook internal sources attracted significantly more shares and comments compared with other QAnon posts. These findings suggest that QAnon pages and groups increasingly isolated themselves from sources outside Facebook while having more internal interactions within the platform, and the endogenous creation and circulation of disinformation might play a significant role in boosting the influence of the misleading narrative within Facebook. The findings imply that the efforts to tackle down disinformation on social media should target not only the cross-platform infiltration of falsehood but also the intra-platform production and propagation of disinformation.

**ACM Reference format:**

## 1 INTRODUCTION

There has been growing concern about the proliferation of disinformation and conspiracy theories on social media [1, 18, 24, 25, 33, 35]. Participating in and endorsing online conspiracy theory discussions is known to lead to rejection of science, unwillingness to involve in prosocial behaviors, and even extreme offline behaviors and violence [38, 40]. Especially, "QAnon," a far-right conspiracy theory based on a set of unsubstantiated claims in favor of the former U.S. President Donald J. Trump, is known to have widely spread on social media platforms since 2017 [16, 61, 70]. QAnon is centered on unsubstantiated and false claims made by an anonymous individual known as "Q," that a global pedophile cabal, including top Democrats, operates a child sex-trafficking ring, and Donald J. Trump would finally unmake and punish its members [51]. It has been reported that Facebook has been particularly plagued with the conspiracy narrative [16, 48], but there has been a lack of systematic investigation of how QAnon propagated on the world's largest social media platform.

This research investigated the propagation of the QAnon narrative based on, to the best of our knowledge, the most extensive dataset about the conspiracy theory on Facebook. Scholars have pointed out the need to understand the ecosystem in which disinformation is produced and propagated within and across online platforms [20, 37, 59, 61, 64, 70]. The QAnon conspiracy theory is known to encourage its audience to assemble and remix different pieces of information [70], so it is even more crucial to examine how online communities discussing QAnon interacted with the broader ecosystem by drawing information from various sources to construct and support their narratives.

Thus, in this work, we examined the activities of Facebook pages and groups that contributed to promoting the conspiracy theory and explored how they constructed the misleading narrative. Specifically, we focused on how Facebook pages and groups used information sources internal or external to Facebook in generating QAnon-related content. This study aimed to answer the following questions: How did the activities of the online groups promoting the QAnon narrative on Facebook change over time? What information sources were used in discussing the conspiracy theory and how the sources of information influenced users' engagement with the social media content? Addressing these questions not only helps understand the information ecosystem of QAnon on Facebook but also sheds light on how misleading narratives are produced and promoted in digital spaces and what factors should be considered in developing strategies to reduce or prevent the spread of disinformation.

This research investigated the propagation of the QAnon narrative based on an extensive dataset about the conspiracy theory on Facebook. Scholars have pointed out the need to understand the ecosystem in which disinformation is produced and propagated within and across online platforms [20, 37, 59, 61, 64, 70]. The QAnon conspiracy theory is known to encourage its audience to assemble and remix different pieces of information [70], so it is even more crucial to examine how online communities discussing QAnon interacted with the broader ecosystem by drawing information from various sources to construct and support their narratives. We expect that exploring the seemingly simple but fundamental questions of the current study would make contribution not only by providing empirical evidence on the structure and dynamics of an online disinformation ecosystem but also by increasing our understanding of the critical socio-technological phenomenon that had profound implications and impacts on society and politics [16, 61].

## 2 RELATED WORKS

### 2.1 Prior Research on Misinformation

There has been growing academic interest in the production and consumption of online misinformation and conspiracy theory content [5, 18, 20–22, 35, 39, 42, 58]. In a study based on a nationally representative sample of Twitter users, Grinberg et al. [20] examined the fake news propagation during the 2016 U.S. Presidential Election and found that most misinformation was shared by a small number of "super-spreaders." Leveraging extensive access to Twitter's historical database, Vosoughi et al. [63] showed that false news stories propagated faster and more broadly than true ones. Starbird et al. [60] investigated URL links shared by Twitter users and found that the majority of the shared domains was alternative media or blogs. They explained that the users attempted to construct conspiracy narratives by referencing these sources. Scholars also examined factors affecting the spread of misinformation [8, 10, 27, 53, 60, 63]. Buchanan [8] found that participants were more likely to share misinformation when they regarded it as true or consistent with their pre-existing attitudes. Considering the platform-level factors, Hussein et al. [27] examined the effect of personalization attributes on the extent of misinformation in future searches and recommendations on YouTube. Focusing on the strategies of conspiracy theorists, Kou et al. [35] identified that Reddit users tried to defend conspiracy theories by selectively citing authoritative information, proposing unknowable risks, casting doubts, and deflecting the burden of proof.

Pioneering studies on QAnon, the conspiracy theory that had immense impacts on American politics and society, are emerging recently [1, 4, 29, 42, 43]. The scholarly effort to understand the actors, the discourse, and the information ecosystem of QAnon has shed light on alternative communication platforms that had helped initiate and grow the conspiracy theory and its influence, and the literature is further expanding to address more complex large-scale issues related to the extreme and misleading narrative. These issues include cross-platform dynamics and inter-platform comparisons regarding the conspiracy theory's evolution and propagation, and the investigation of major digital platforms, such as Facebook and Twitter, focusing on how these giant platforms contributed to and were affected by QAnon.

Regarding alternative communication platforms, Papasavva et al. [42] explored the activities of QAnon communities on Voat.co, an alt-tech news aggregator, and showed that the conspiracy narrative had close relationships with so-called "Q drops" and other conspiracy theories. Scholars have also made considerable progress in understanding Parler, a fringe social networking platform that played a critical role in the evolution of right-wing extremism and conspiracy theories. Based on an extensive dataset obtained from the website, Bär and the coauthors [4] presented crucial observations about QAnon supporters: about 5.5% of Parler users were QAnon supporters, and these users, compared with non-QAnon supporters, created more content and maintained larger social networks, which enabled the QAnon supporters to promote their narratives faster and further. Jakubik et al. [29] argued that, while Trump supporters on Parler reacted to the storming of the U.S. Capitol with increased negative sentiment, QAnon supporters did not display significant changes in their sentiment.

Inter-platform perspectives are essential in understanding how disinformation and conspiracy theories evolve in the broader information ecosystem beyond the boundaries of platforms. One of the most extensive and rigorous studies on this issue was conducted by Aliapoulios et al. [1]. Based on multi-platform datasets including 4chan, 8kun, Reddit, Voat, and other aggregation sites, the researchers found that, although the QAnon narrative's initial dissemination was limited to a few sub-communities, the narrative persisted even after its being banned from some platforms. The research also showed that Q drop messages themselves were not openly toxic or threatening, but the interpretation of these messages within conspiracy theory communities tended to weaponize the narrative and instigate violence. Based on the investigations of Italian QAnon supporters on Twitter, Telegram, and Facebook, Pasquetto et al. [43] explored how they constructed an "infrastructure of disinformation" across different digital communication platforms.

Despite the collective academic effort, there has been a gap in understanding how online communities on Facebook contributed to developing and spreading the conspiracy theory. Thus, advancing the recent line of studies, the present work aims to provide a useful and crucial "building block" in understanding the information ecosystem of disinformation by exploring the existence of QAnon conspiracy theory in the world's largest social media platform, by investigating how Facebook pages and groups promoting the rhetoric of QAnon were connected with the broader information ecosystem, and by examining how these connections changed over time and were associated with the engagement of Facebook's user base.

## 2.2   Misinformation and Conspiracy Theory on Facebook

Facebook has been criticized for failing to suppress misleading and unsubstantiated narratives on its platform [14, 16, 30]. It has been suspected that Facebook pages and groups have produced and promoted conspiracy theories, unverified rumors, and inaccurate information [7, 10, 30, 31, 54]. However, most of the past research focused on Twitter and other platforms, as discussed above. Despite its importance as the world's biggest social media, the number of empirical investigations of Facebook is still insufficient, and more importantly, research on how disinformation and conspiracy theories are discussed and exchanged on Facebook is lacking.

Past studies about misinformation and conspiracy theories on Facebook have been based on (1) Facebook's API, which is not widely used by researchers currently due to its limitations, (2) access to Facebook's database as an insider of the company, which is not a feasible option for academic researchers, or (3) manual data collection, which is not an optimal solution for large-scale quantitative investigations [5, 10, 17]. For example, Friggeri et al. [17] examined how fact-checking affected rumor propagation. Del Vicario et al. [10] and Bessi et al. [5] compared Facebook pages posting science news and those disseminating conspiracy theories. Johnson et al. [32] analyzed the patterns of Facebook pages with pro- and anti-vaccination viewpoints and argued that the growth of anti-vaccination pages was faster than the pro-vaccination pages. Recently, researchers started paying attention to CrowdTangle, a public insights tool owned and operated by Facebook, for Facebook data collection (e.g. Yang et al. [68]).

The present study focuses on examining the use of different information sources in promoting disinformation and its impact on user engagement, which could shed light on the ecosystem of online misleading narratives and on how these narratives are amplified in digital spaces. A few previous investigations provide findings that are relevant to the current research. Bessi et al. [5] reported that conspiracy content about scientific issues received more likes and shares on Facebook compared with mainstream news content. Mocanu et al. [39] studied how alternative news, mainstream news, and self-organized online political movements were associated with user engagement on Facebook. They discovered that those who preferred alternative information sources were more susceptible to misinformation.

## 3    METHOD AND DATA

In the present study, QAnon posts were defined as Facebook posts containing terms and slogans used in the context of the QAnon conspiracy theory, and QAnon clusters were defined as public Facebook pages and groups that created multiple QAnon posts. QAnon clusters contributed to the spread of the conspiracy narrative by repeating the terms and phrases related to QAnon in the posts they generated.

In total, we identified 2,813 QAnon clusters that were active during the period between November 1st, 2017, and November 30th, 2020, using a multi-staged data collection method that enabled us to obtain a comprehensive Facebook dataset. These QAnon clusters created 117,553 QAnon posts, and Facebook users following the clusters received QAnon posts in their social media timelines over 3.1 billion times in total (See Appendix A.2 in Supplementary Information for details). Example QAnon posts are shown in Table 1.

Table 1. Examples of QAnon Posts

| Example | Text content |
|---|---|
| 1 | "#PatriotsUnite Stand together! Where we go One We go All … #WWG1WGA" |
| 2 | "TRUMP SENDS OUT THE STORM TROOPERS … @realDonaldTrump #WWG1WGA" |
| 3 | "… CNN to frame upcoming Pizzagate arrests of politician pedophiles … #Pedophiles #DCElites" |
| 4 | "… We want America First, We do not want a world bank .Hey Schummer ,Pillosi ,Warren Sanders, Watters, Gram , ,Mconnell, and all the other NWO (New World Order ) puppets in the Democrat and Republican Parties … George Sorros should be investigated by the "Federal Government" #maga" |
| 5 | "… the Glory awaiting humanity! … #RedeemingTheFall #WWG1WGA" |

*Note.* The text was partly removed to reduce the possibility of the identification of specific communities and users. Removed parts were replaced with "…".

### 3.1 Data Collection

*3.1.1   Data collection method.*    To collect data on QAnon clusters, we implemented a multi-staged data collection method combining data retrieval via CrowdTangle and computational techniques to discover new Facebook accounts. CrowdTangle (CT) is a public insights tool owned and operated by Facebook. Researchers and journalists using CT can search and download data on public Facebook pages and groups. For researchers, CT offers two major advantages. First, it provides a powerful search function for Facebook content. CT users can search for Facebook posts that contain specific keywords in the text and images included in the posts. Second, CT users can access the entire historical record of Facebook accounts tracked by CT. CT tracks over 7 million influential Facebook pages, groups, and verified accounts but does not track all public Facebook accounts, which limits the scope of Facebook data retrieved from CT [9, 65]. Thus, we implemented a new data collection method to overcome the limitation.

CT allows users to add and track new Facebook accounts which do not exist in its system [65]. Based on this feature, we designed a "snowball" data collection method that iterates data downloading, link analysis, and data request, which enabled us to obtain sizeable data beyond the limited scope of the CT database. (The Facebook accounts we identified during data collection are now on the CT database and accessible to other CT users.)

The data collection started on December 24th, 2020, and was completed on January 8th, 2021. The timing of data collection ensured that the collected data was not significantly influenced by Facebook's platform-level takedown of content created by Trump supporters and QAnon theorists attempting to delegitimize the outcome of the 2020 Presidential Election [28].
This study was approved by the Institutional Review Board of Stanford University.

*3.1.2 Keyword selection.* The QAnon conspiracy theory has been built upon and closely related to other conspiracy theories, such as the "pizzagate" conspiracy theory and the "deep state" conspiracy theory [51]. The QAnon narrative is based on its own characteristic slogans and dialects, such as "qarmy," but it also adopts terms from the related conspiracy theories, such as "pedogate." Thus, we developed two sets of keywords to capture not only Facebook pages and groups relying on QAnon-specific phrases but also those referencing the QAnon context in its broad relationship with other related conspiracy theories.
We defined the two sets of keywords by expanding and modifying keywords selected in previous studies [15, 62]. First, the *core* set of keywords consists of 14 terms used specifically in the QAnon context. Second, the *extended* set of keywords includes all terms in the core set and additional terms used broadly in the QAnon and related conspiracy context (40 terms in total). The data collection process aimed to identify Facebook pages and groups which mentioned one or more core keywords and multiple extended keywords across multiple posts, as described in 3.1.3 in detail.

1. *Core* keywords: "qanon," "qarmy", "weareq," "weareallq," "17anon," "qdrop," "qpost," "qsentme," "qproof," "asktheq," "greatawakening," "wwg1wga," "thestorm," and "theperfectstorm" (14 terms).
2. *Extended* keywords: "qanon," "qarmy", "weareq," "weareallq," "17anon," "qdrop," "qpost," "qsentme," "qproof," "asktheq," "greatawakening," "wwg1wga," "thestorm," and "theperfectstorm," "painiscoming," "darktolight," "enjoytheshow," "taketheoath," "trusttheplan," "questioneverything," "draintheswamp," "sheepnomore," "takebackcontrol," "digitalsoldiers," "wearethenews," "truthseeker," "followthewhiterabbit," "savethechildren," "saveourchildren," "deepstate," "pedowood," "pedogate," "pedovores," "pizzagate," "adrenochrome," "spiritcooking," "thepeoplearesick," "epsteindidntkillhimself," "childtrafficking," and "humantrafficking" (40 terms).

Although the keywords used in this research do not constitute an exhaustive list of all words and phrases representing the QAnon narrative, the selection of keywords listed above and the filtering rules explained in 3.1.3 considered both the key characteristics of the QAnon narrative and its relationships with other conspiracy theories, adopting and extending previous research on this issue [1, 15, 42, 62].

*3.1.3 Iterative data collection process.* We conducted three rounds of data collection. In the first round, we used CT's Post Search API (Application Programming Interface) to download all English posts created between January 1st, 2010 and December 23rd, 2020 that contained at least one keyword in the extended set. Based on the downloaded post data, we created a list of Facebook pages and groups that (i) generated at least 5 posts in the search results, (ii) stated at least 5 different extended keywords in their posts appearing in the search results, and (iii) stated at least one core keyword in their posts appearing in the search results. These three criteria ensure that selected pages and groups repeatedly created content discussing details of the QAnon narrative. For example, pages and groups that uploaded few QAnon-related news articles cannot satisfy the first and second criteria. Pages operated by legitimate news media, such as the Fox News page, are unlikely to satisfy the second criteria. The accounts that passed the filtering process of the first round were called the "first-round accounts," and their posts included in the search results were called the "first-round posts."

In the second round of data collection, we started by analyzing all Facebook internal URLs included in the first-round posts and discovering new Facebook accounts that were linked with the first-round accounts via these URLs. We then requested CT to retrieve data from these newly identified Facebook accounts. After that, we utilized the Post Search API to retrieve all English posts that were generated by these new accounts between January $1^{st}$, 2010 and December $23^{rd}$, 2020, and contained at least one keyword in the extended set. Based on the downloaded data, we created a list of Facebook accounts satisfying the same three conditions used in the first round. The accounts that passed the filtering process of the second round were called the "second-round accounts," and their posts included in the search results were called the "second-round posts."

The third round started by analyzing all Facebook internal URLs included in the second-round posts and identifying new accounts that were linked with the second-round accounts via URLs but were not included in the first-round and second-round accounts. We then requested CT to retrieve data on these newly identified Facebook accounts and retrieved all English posts which were created by the accounts between January $1^{st}$, 2010 and December $23^{rd}$, 2020, and contained at least one keyword in the extended set. From the downloaded content data, we identified Facebook pages and groups satisfying the same three conditions for the QAnon account selection. Only one new QAnon cluster was identified in the third round, and thus we stopped the iterative data collection process after finishing this round.

Lastly, all post data retrieved during the three rounds were combined, and then posts created before November $1^{st}$, 2017, or after November $30^{th}$, 2020 were removed from the combined data, since the QAnon conspiracy theory allegedly started in October 2017.

It is known that QAnon supporters hijacked some of the slogans that had long been used by human rights and anti-trafficking groups [50]. Therefore, we conducted additional data cleaning by reviewing 39 Facebook pages and groups in the collected data that passed the filtering with only 5 keywords: "qanon," "savethechildren," "saveourchildren," "childtrafficking," and "humantrafficking." Two researchers reviewed these pages and groups and identified that 38 of them were pure human rights and anti-trafficking communities. These 38 pages and groups could pass our keyword filtering criteria because (i) they had been using the anti-trafficking slogans related to human rights purposes and (ii) they had stated "qanon" while denying their relationships with or expressing their concern about the conspiracy theory. We removed the 38 pages and their posts from the collected data.

The resulting data contained 117,553 QAnon posts generated by 2,813 QAnon clusters (2,070 Facebook pages, 743 Facebook groups). Although, without access to Facebook's internal knowledge about its website architecture and database structure, we cannot assert that the collected data include a complete set of Facebook pages and groups satisfying our search criteria, this method enabled researchers to observe an extensive number of accounts beyond the limited pool offered by CrowdTangle, one of the very few ways allowed for quantitative researchers investigating Facebook.

One may point out that discovering new accounts based on Facebook internal URLs might contribute to the overrepresentation of internally shared content. However, it should be noted that the second and third rounds retrieved "all content" in the pages and groups discovered in the previous round, and thus the retrieved content in these rounds also includes posts not shared by any other pages and groups. Using Facebook internal URLs as a clue to discover hidden accounts allowed us a more complete observation, and we do not think that the retrieval process itself imposed a meaningful penalty on unshared content. Also, the present research focuses more on over-time variations in Facebook internal and external sources, rather than making conclusions based on a direct comparison between different information sources. Ultimately, gauging the level of potential bias and its potential influences must be a challenge facing all external researchers who cannot obtain information about the population of platform users or unbiased samples from it. The data retrieval method laid out here can be viewed as a way to bypass restrictions imposed by Facebook and to produce comprehensive datasets despite the practical barriers.

## 3.2 Preliminary Classification of QAnon clusters

To understand the topics generally discussed in QAnon clusters, we analyzed the top 200 clusters based on the number of QAnon posts generated in the given time period. Excluding 46 clusters that were not accessible as of April 2021, two subject-matter experts manually accessed and reviewed the QAnon cluster webpages and identified 7 different types of QAnon clusters. These types include supporters of politicians or journalists, general discussion groups, human rights groups, conspiracy theory groups, and spiritual discussion groups. Example QAnon clusters of each type are shown in Table 2.

Table 2. Preliminary Classification of QAnon Clusters

| Category | N | Example page/group names |
|---|---|---|
| Supporters of politicians and journalists | 51 | "… Trump Only …", "… Keep America Great …" |
| Independent news organization | 24 | "…News and Views" |
| Forums (Groups sharing information about general issues) | 24 | "… Truth Movement…" |
| Human rights group and non-profit org | 19 | "… Family Rights", "… Against Trafficking", "Saveourchildren…" |
| Conspiracy theory group | 18 | "… Truth Seekers", "Exposing the Lies of …" |
| Spiritual discussion group | 8 | "… Cosmic News", "Galactic …" |
| Miscellaneous | 10 | |

*Note.* Names were partly removed to reduce the possibility of the identification of specific pages and groups. Removed parts were replaced with "…". The top 200 clusters based on the number of QAnon posts were chosen for this preliminary classification, and among them, 46 clusters that were not accessible as of April 2021 were excluded.

## 3.3 URLs and Information Sources

We explored the information ecosystem that enabled QAnon clusters by analyzing URLs (Unified Resource Locators) included in QAnon posts and identifying information sources used by the clusters. Facebook pages and groups tend to rely on multiple information sources in generating their content. Information sources used in a Facebook post can be traced back by analyzing URLs included in the post. Each URL corresponds to a source from which a piece of information or content was drawn, such as a photo included in a post, a newspaper article linked in a post, and a website address referenced in a post. A post can include multiple URLs, and CrowdTangle provides information on all URLs included in a post. For example, if a post created by Cluster A includes a hyperlink to a Fox News Article while resharing another post from Cluster B, and if the post from B included a video uploaded by B while mentioning a CNN article by including a hyperlink to it, the data on the Cluster A's post will include the URL of the Fox News article, the URL of the photo uploaded by Cluster B, and the URL of the CNN article. At the web domain level, the post in Cluster A is drawing information from three different sources: foxnews.com, facebook.com, and cnn.com.

Facebook users often use link shortening services to create a new URL and include it in their social media content instead of its original full URL. Link shortening services remove some of the information available in a full URL while creating a shorter version of it. For example, bit.ly, a URL shortening service, shortens https://en.wikipedia.org/wiki/QAnon into https://bit.ly/3hKCAOQ. In this example, while the original full URL shows the actual domain (wikipedia.org), its shortened version does not. Thus, we identified shortened URLs included in Facebook posts, retrieved their full URLs, and identified their actual domains. To identify shortened URLs, we built a list of 55 popular link shortening services by supplementing the list of services reported by Yang et al. [68] and adding 6 more services to the list: fb.me, is.gd, chng.it, tobtr.com, eepurl.com, cutt.ly, and gf.me. The full list of 55 shortening services covers all shortening services responsible for at least 30 URLs in our dataset. Based on the list of link shortening services, we detected shortened URLs in Facebook posts and identified their original URLs and actual domains. For shortened URLs whose original URLs were not identifiable, domains of their link shortening services were considered as their actual domains. The proportion of unidentifiable URLs was minimal: even the most frequent case, t.co, accounted for only 0.01% of all URLs.

## 3.4 Classification of URLs

We identified top-level domains of all URLs. For example, the domain of an URL, www.nytimes.com/article/what-is-qanon.html, is nytimes.com. Each URL was then categorized into one of the two source types according to its domain.

*3.4.1 Facebook internal sources.* URLs whose domains are facebook.com, fb.com, and fb.watch were categorized as Facebook internal sources. The existence of Facebook internal sources in a post indicates that the posting user included or referenced Facebook in-house content (e.g., photos, videos, and notes stored on the platform), accounts (e.g., pages, groups, and individual profiles), or services.

*3.4.2 External sources.* URLs that are not Facebook internal sources were categorized as "external sources." Three subgroups of external sources were further identified: social media sources, news sources, and low credibility sources, which are known to be critical in understanding the creation and propagation of disinformation [3, 20, 22, 67].

URLs whose domains were included in the social media domain group were categorized as "social media sources." The social media domain group consists of the domains of social media services that were included in Pew Research Center's reports on social media from 2012 to 2020, except Facebook [11–13, 19, 26, 45, 55–57]. This group consisted of the domains of Instagram, Reddit, Twitter, YouTube, TikTok, Tumblr, Google+, Twitch, Vine, WhatsApp, and Snapchat. (For Google+, URLs had to match with plus.google.com at the subdomain level. For other social media services, URLs were matched at the top-level domain, e.g., twitter.com and instagram.com.)

URLs whose domains were included in the news domain group were categorized as "news sources." The news domain group includes domains that were included in the following lists: "Hard news domains" on Bakshy et al. [3], "News media sites" on Yang et al. [69], "Green" and "Yellow" domains on Grinberg et al. [20], and "Newspapers" and "Digital-native news outlets" labeled by Pew Research Center [47]. Domains in the low credibility domain group were excluded from the news domain group. This group consists of 529 domains.

URLs whose domains were included in the low credibility domain group were categorized as "low credibility sources." The low credibility domain group included domains categorized as "Black," "Red," or "Orange" sources in Grinberg et al. [20], which extended previous lists of fake news sources curated by Guess et al. [23], Allcott and Gentzkow [2], Snopes.com, and Buzzfeed; and domains identified as "very low credibility" and "low credibility" sites by Media Bias and Fact Check (mediabiasfactcheck.com). The number of domains in this group was 889.

## 4 RESULTS AND DISCUSSION

### 4.1 The Activity of QAnon Clusters

A descriptive overview showed that the activity of QAnon clusters included in the current dataset had significantly intensified in 2020. First, the number of active clusters in our data drastically increased since 2017 (Fig. 1A). Second, active clusters in 2020 generated 2.0 and 1.9 times more posts than those in 2019 and 2018, respectively (Fig. 1B). Third, the slope of the cumulative curve and the distribution of colors in Fig. 1C demonstrate a surge in the number of clusters that newly participated in 2020 and intense activities of these new clusters. Specifically, when all clusters were classified into four different cohorts based on the year that their first QAnon post was created, the 2020 cohort produced 7.1 QAnon posts per week on average, which was 8.4 and 20.5 times greater than the 2019 and 2018 cohorts, respectively (Fig. 1D). Lastly, the number of posts generated within the first month was the highest for the 2020 cohort, and it was 2.1 and 3.6 times greater than the 2019 and 2018 cohorts, respectively (Fig. 1E). See Appendix A for detailed analyses.

It is worth emphasizing that we report these observations to provide preliminary descriptive information about the main dataset of the current research, before discussing detailed patterns and associations within the dataset in Sections 4.2 and 4.3. Although we believe that these descriptive findings are informative, the primary goal of this research is not to claim that the problematic

Facebook QAnon content showed an exclusive pattern of a surge in the entire platform or to reveal what caused the explosive growth. Despite social media's significant impacts on society and the need for thorough academic investigations, complete validation of these hypothetical claims is deemed not very feasible at the moment, largely because of the limited scope and scale of Facebook data that are accessible to external researchers.

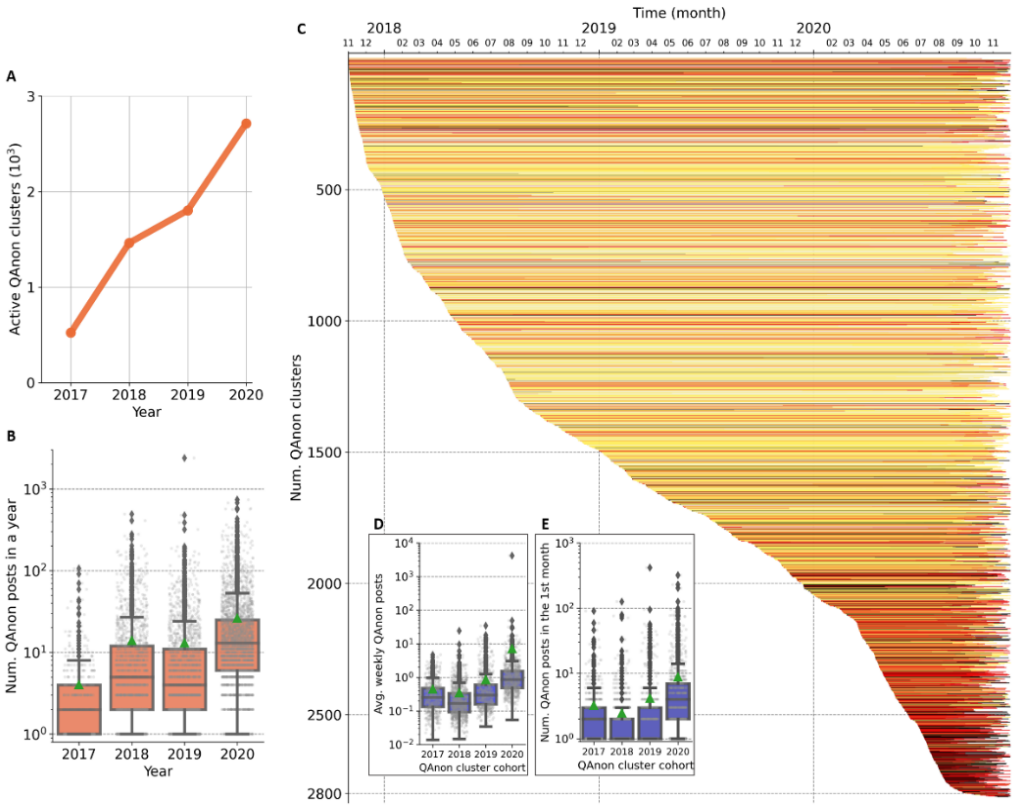

Fig. 1. The activity of QAnon clusters on Facebook. **A**. The number of active clusters. Active clusters are defined as clusters that created at least one post in a certain time period. A dot indicates the number of active clusters in a given year. **B**. The number of posts generated by an active cluster each year. A green triangle indicates a mean. Each grey dot represents a cluster. **C**. Cluster lifetime and the average number of weekly posts. The lifetime of a cluster is defined as the time duration between its first and last QAnon posts. Each line corresponds to a cluster, starting at the date of its first post and ending at the date of its last post. The color of a line represents the average number of posts per week created by the cluster, $N_w$. Black: $N_w > 1.5$, Brown: $N_w > 1.0$, Red: $N_w > 0.5$, Yellow: $N_w > 0.1$, Grey: $N_w \leq 0.1$. **D**. The average weekly posts generated by a cluster during its lifetime. Clusters were grouped into four different cohorts based on the year that their first QAnon post was generated. **E**. The number of posts generated by a cluster in its first month.

However, due to potential social and policy implications that these surging patterns have, one may still question if the increasing volume of Facebook content over time shown in this study is a general trend that Facebook posts containing any keywords show in the given period. Another potential argument is that the volume of Facebook content containing any keywords had increased since the beginning of the pandemic (e.g., due to the increased screen time during lockdowns). We attempted to explore these claims by checking if the number of posts containing other keywords shows a longitudinal pattern similar to those containing "qanon." We collected data on various topics using CT's Post Search API to retrieve content from public Facebook pages and groups and counted the monthly number of all posts containing each of the following keywords: keywords representing relatively general political issues ("gun control," "immigration ban"), political figures ("barack obama," "hillary clinton," "donald trump"), pandemic-related terms ("covid-19," "social distancing"), non-political issues related to conspiracy theories ("flat earth," "chemtrail"), and general non-political topics ("weather," "dinner recipe," "beyonce," and "basketball"). Extracting data on various topics and comparing their temporal patterns with QAnon was made possible thanks to the keyword searching method, which enabled the extensive validation without the considerable resources and time required for the iterative processes for the main data collection. For each keyword, the normalized number of Facebook posts was computed for each month by dividing the number of all Facebook posts containing a keyword in a given month with the 107-month average number of posts containing the keyword. As shown in Fig. 2, the additional topics tested here do not display a pattern that shows a consistent surge throughout 2020. Although the two keywords related to the COVID-19 pandemic, "covid-19", and "social distancing", showed a surge that peaked in early 2020, their monthly volume decreased later that year, creating a temporal pattern different from that of QAnon. For other keywords tested in this examination, noticeable surges in recent years were not detected.

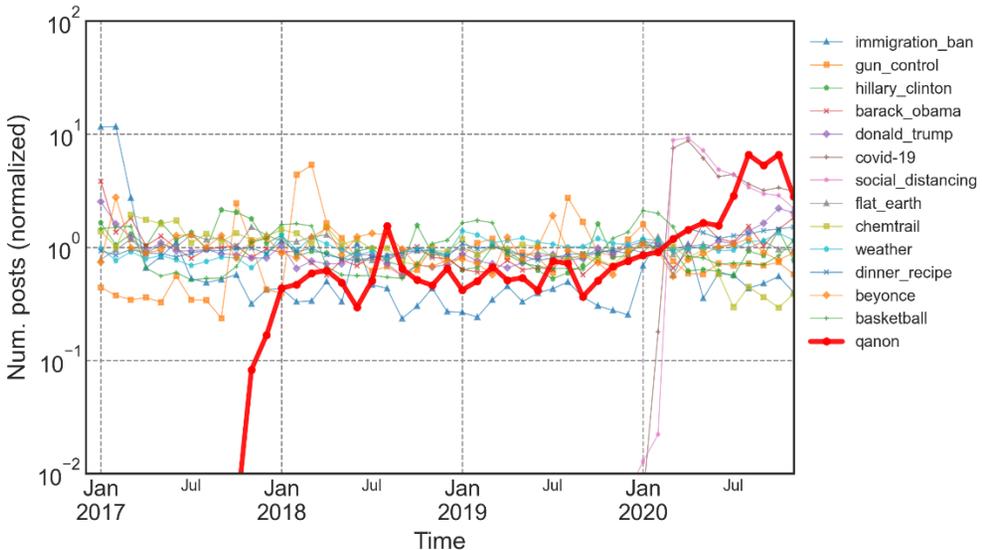

Fig. 2. The volume of Facebook content containing a specific keyword

## 4.2 Information Sources for the QAnon Narrative

To investigate the information ecosystem that enabled the production and distribution of the QAnon narrative, we analyzed information sources used by QAnon clusters.

We examined the proportion of posts using information sources within the Facebook platform (i.e., Facebook internal sources) and other sources outside the platform (i.e., external sources). The prevalence of posts using Facebook internal sources had increased significantly each year from 30.0% in 2017 to 50.0% in 2020, while the use of external sources steadily decreased over time from 74.4% in 2017 to 52.7% in 2020. These results show that the misleading narrative increasingly more relied on information sources within Facebook, drawing less and less information from outside. It is possible that the increasing separation is a platform-level shift, not limited to the QAnon narrative, but this observation still has significant implications, regarding the role played by the platform and potential interventions to reduce the conspiracy theory narrative. Fig. 3C and D visualize the drastic decrease and increase in detail. These changes are discussed in more detail in Appendix B.2-3.

We further investigated three subcategories of external sources: social media sources, news sources, and low credibility sources (Fig. 3B). The proportion of posts using social media sources sharply declined from 40.6% in 2017 to 24.0% in 2020. It was also found that less than 5% of posts in a cluster referenced low credibility sources, and the proportion even decreased each year from 5.0% to 1.3% between 2017 and 2020. The proportion of posts utilizing news sources ranged between 6 and 9%. Detailed numeric statistics of subcategories are provided in Appendix B.4-6.

These patterns exhibit interesting similarities and differences to Facebook discourses on other contentious issues, such as vaccines, climate change, and Black Lives Matter [33, 34]. Studies based on a similar iterative multi-staged data collection method using CT reported that, while the long-term increase in the proportion of Facebook internal sources was also observed among Facebook groups and pages discussing vaccines (among both pro- and anti-vaccine groups) and climate change (among both groups denying and advocating climate change) [33, 34], it was not observed among Facebook pages and groups supporting Black Lives Matter or its counter-movements [33]. Each external source subtype's longitudinal change presented here is also distinguishable from what has been discovered from Facebook pages and groups discussing vaccines [34]. Considering these differences, interpreting the over-time changes depicted in this study as direct consequences of Facebook's platform-wide policy changes or their effort to combat misinformation has a possibility of overgeneralization and thus should be done with caution.

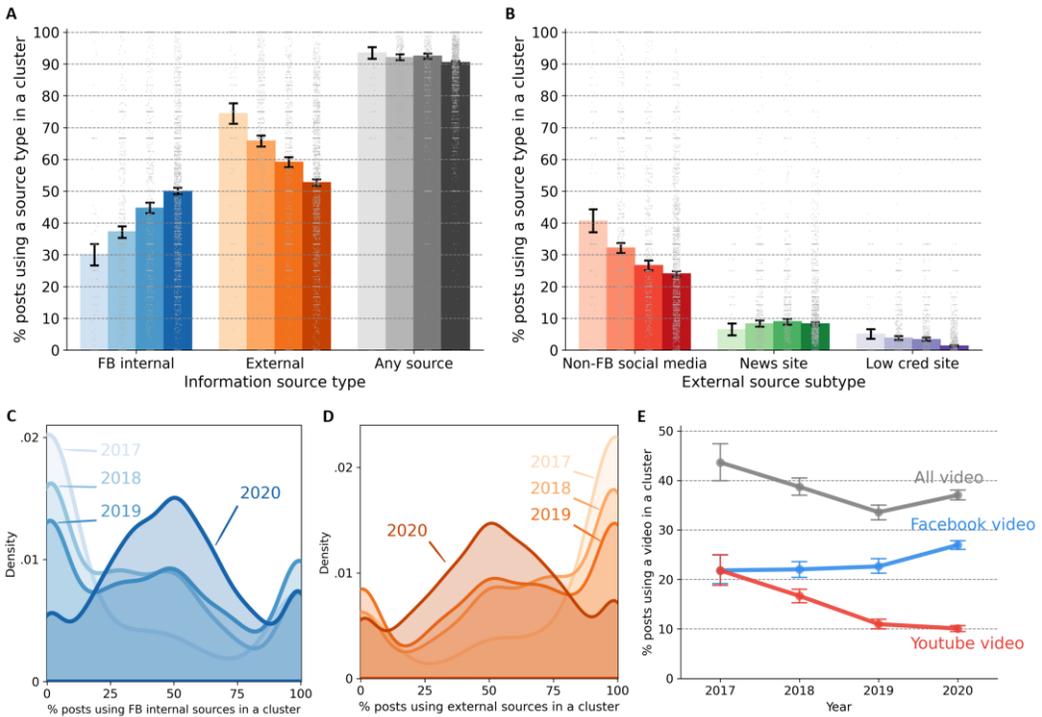

Fig. 3. Information sources for QAnon clusters. **A**. The average proportion of posts in a cluster using different types of sources. "Any source" refers to the proportion of posts in a cluster containing one or more URLs. Bars and error bars indicate mean ± 95% confidence interval. Each gray dot represents an active cluster in a given year. Four bars in each group represent different years: 2017 (the leftmost), 2018, 2019, and 2020 (the rightmost), respectively. **B**. The average proportion of posts in a cluster using different subtypes of external sources. **C**. The distribution of proportions of posts in a cluster using Facebook internal sources. **D**. Posts in a cluster using external sources. **E**. The average proportion of posts in a cluster using the two different video sources. Error bars indicate mean ± 95% confidence interval.

We compared Facebook internal sources with the most dominant external source of QAnon clusters, YouTube (Detailed information is available in Appendix B.7). Consistent with the aforementioned trends, the proportion of posts using YouTube videos in a cluster has decreased between 2017 and 2020, in contrast to the increase in the prevalence of posts using Facebook videos over time (Fig. 3E).

## 4.3 Association between Information Source Use and User Engagement

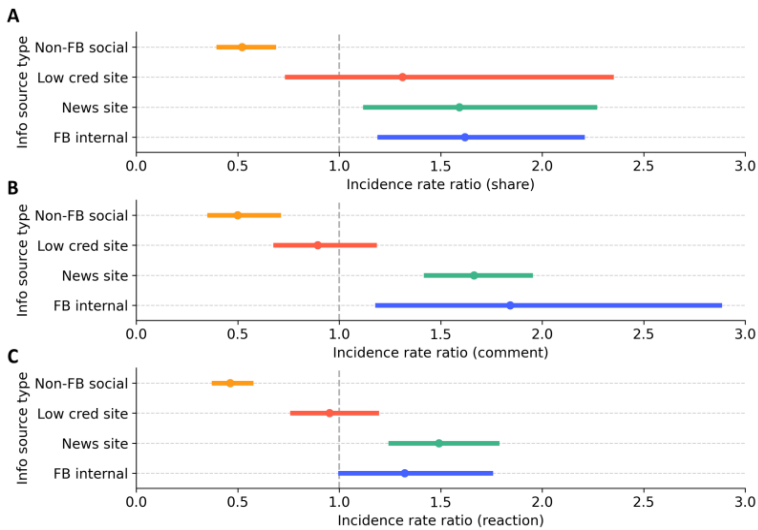

Fig. 4. Information sources and QAnon content engagement. Incidence rate ratios (IRRs) were calculated using the results of regression models fit to the number of shares, comments, and reactions of a post, respectively, controlling for covariates. All standard errors were clustered at the QAnon cluster level, and all models were estimated with cluster-robust standard error at the QAnon cluster level. The dots and lines represent estimated mean IRRs ± 95% confidence intervals. **A.** Associations between the inclusion of each information source type in a post and the number of shares of the post. **B.** Information source types and the number of comments. **C.** Information source types and the number of reactions.

We investigated whether the use of a certain type of information source in a post was positively or negatively associated with its user engagement. We found that QAnon posts using Facebook internal sources produced more shares and comments than those not using internal sources, while posts with external sources induced fewer shares, comments, and reactions than posts without external sources (Appendix C.1).

When the three subtypes of external sources were considered (Fig. 4), QAnon posts with news sources received more shares (IRR = 1.59, 95% CI [1.12, 2.27], P = .010), comments (IRR = 1.66, 95% CI [1.42, 1.95], P < .001), and reactions (IRR = 1.49, 95% CI [1.24, 1.79], P < .001) compared with those not including news sources. Also, QAnon posts using Facebook internal sources appeared to attract more shares (IRR = 1.62, 95% CI [1.19, 2.21], P = .002) and comments (IRR = 1.84, 95% CI [1.18, 2.89], P = .008) than other QAnon posts. These results show that (1) referencing legitimate news sources within the conspiracy theory context on Facebook (i.e., drawing content via URLs from the news sites to Facebook QAnon posts and their misleading narrative), and (2) using Facebook internal sources in producing and supporting the conspiracy theory content (i.e., drawing content via URLs from the internal sources to Facebook QAnon posts and their misleading narrative) were associated with greater attention and reactions from users. On the other hand, the inclusion of non-Facebook social media sources was negatively associated with shares (IRR = 0.52, 95% CI [0.39, 0.69], P < .001), comments (IRR = 0.50, 95% CI [0.35, 0.71], P < .001), and reactions (IRR = 0.46, 95% CI [0.37, 0.58], P < .001). (The negative associations could be because social media links routed Facebook users to other platforms, reducing their engagements with Facebook posts as a result. Examining potential mechanisms like this one is, however, outside the scope of this research and also beyond the capabilities of the current data.) See Appendix C.2 for the detailed model specification and results.

The significant associations between user engagement, and Facebook internal sources and non-Facebook social media sources were still identified in a different statistical model chosen for a robustness check (Appendix C.3). Evidence supporting significant associations between the inclusion of low credibility sources and user engagement measures was not found in this analysis. It is also worth mentioning that we are not making claims on causal relationships or causal mechanisms in these analyses. For example, an association can show that "posts including news sources received more shares," but saying "including news sources led to the difference in shares" based on the association inevitably accompanies certain levels of speculation.

The positive association between user engagement and the use of Facebook internal sources appears to be more consistent in the context of QAnon than other issues, such as vaccines [34]. The insignificant association between user engagement and the use of low credibility sources is also distinctive, compared with a past finding that vaccine posts including low credibility sources were significantly associated with more shares, comments, and reactions [34]. The negative associations between user engagement and the use of non-Facebook social media sources, on the other hand, are aligned with the findings from the vaccine context [34].

## 5   DISCUSSION AND CONCLUSIONS

Despite significant concerns about disinformation and conspiracy theories encroaching the world's largest social media, the lack of systematic research has prevented us from understanding the status and developing appropriate strategies against the social problem. The current study examined the activities and the information ecosystem of online communities that contributed to promoting the QAnon narrative, one of the most influential conspiracy theories in the past decade. The activities of QAnon pages and groups in the present dataset significantly surged in the year of the 2020 Presidential Election. By analyzing their use of information sources, we further discovered the following.

First, QAnon pages and groups increasingly relied on resources within Facebook and disconnected themselves from the ecosystem outside the platform. It suggests that QAnon's narrative proliferated within their self-sustaining Facebook environment which is separated from other platforms. Because QAnon posts based on internal sources induced higher levels of user engagement, this trend might also help boost the influence of the QAnon narrative on Facebook. The decrease in the inter-platform connections identified in this research is a long-term trend spanning multiple years. Hence, it does not necessarily conflict with past studies focusing on shorter time periods and showing that QAnon communities on Facebook maintained active connections with the broader information ecosystem [43]. Despite its long-term decline, cross-platform information flows could be still active and meaningful, and the present findings do not negate their importance and influence.

Here we can speculate about two potential factors that might contribute to and interact with the increasing intra-platform information flow. From the user perspective, the increasing platform-level isolation may indicate that the growth of the QAnon narrative on Facebook had been driven by strong network effects [24, 52]. Facebook's huge user base might initially attract QAnon promoters whose goal was to persuade and recruit more supporters. As Facebook QAnon communities grow, rather than bringing in and aggregating information from outside the platform and becoming a "hub" or "gateway" site of QAnon discourse, they might choose to ramp up and diversify social interactions within the platform, which, in turn, might successfully lead to the growth of Facebook QAnon communities. This cycle of positive reinforcement is especially likely given that it has been the central driver of the explosive growth and continuing dominance of the social media platform itself [66]. From the platform perspective, on the other hand, the increasing significance of intra-platform processes may represent the effects of the platform's recommendation algorithms. The algorithms are designed to stimulate and facilitate social interactions and information exchange within the platform by recommending new social connections or new information that is highly likely to generate reactions or retransmissions [36]. Thus, algorithms facilitating the interactions and connections among QAnon pages and groups and suggesting in-house content and user-generated materials to the communities might contribute to the proliferation of endogenous processes and the separation of the platform from the broader information ecosystem.

Second, the results imply that combining legitimate news media sources with misguiding narratives to create "alternative facts" [48] might effectively stimulate public attention and reaction to conspiracy theory content [58, 70]. Lastly, previous studies reported that the prevalence of information from low credibility sources is generally limited [20, 22], and the current study reveals that it still holds even among communities on social media discussing conspiracy theories.

The present research is not without limitations. Although we could identify a large number of public pages and groups discussing QAnon, data on accounts and content removed by Facebook, and information about private accounts and individual profiles are not available to researchers. Especially, because Facebook had allegedly deleted QAnon posts, pages, and groups since August 2021 [16], the current dataset might overrepresent QAnon content and clusters that had not been removed by Facebook's algorithms and content moderation. More industry-academia collaboration and data sharing for research purposes would be essential in examining the role played by algorithms and platform-wide policies in the spread of disinformation. Also, generalizing the findings of this research should be done with caution because, despite its scale and scope, the current dataset is not a representative sample of the entire platform. Researchers should also be careful in expanding the implications of the present study to other misleading narratives and other social media services. Lastly, although the researchers manually examined a subset of the identified QAnon-related accounts and performed a qualitative categorization to ensure the accuracy of data collection and to better understand the context, we expect that more rigorous quantitative assessment combining human coding and machine classification would help guarantee data quality and accuracy in future research.

Despite the limitations, this study provides empirical evidence urgently needed to understand the critical social phenomenon and to consider potential remedies for disinformation that propagates on social media. The findings imply that tackling the intra-platform creation and circulation of disinformation within social media platforms could be imperative in curbing infodemics. This study also indicates that strategies focusing only on the potential intrusion of content from suspicious external sources may not be sufficient nor efficient in minimizing the impacts of disinformation. The increasing importance of the endogenous processes within communication platforms and the disproportional impacts that internally forged false narratives might have on users suggest that social media platforms have the potential to become a giant incubator and amplifier of disinformation, boosting the reach and the impacts of disinformation on a large number of people. Aligned with the reports on Facebook's alleged failure to suppress the explosive growth of disinformation [16], this study suggests that more independent and external investigation and monitoring of the communication platforms is essential in not only increasing our knowledge of disinformation but also developing more effective and transparent solutions for the social problem [41].

**ACKNOWLEDGMENTS**

## A   Appendix A: The Activity of QAnon Clusters

*A.1   The number of active clusters.*   Active clusters are defined as clusters that created at least one QAnon post in a certain time period. The number of active clusters in 2017 was 524, and it increased to 1,464, 1,804, and 2,714 in 2018, 2019, and 2020, respectively (Fig. 1A). Monthly, the number of active clusters was 367 in November 2017 and steadily increased, peaking at 2,131 in August 2020, and then decreased afterward. The number of active clusters was 1,029 in November 2020. This pattern might reflect the fact that Facebook started cracking down accounts spreading QAnon and their posts in August 2020 [16].

*A.2   The number of users who received QAnon posts directly.*   The number of users who received QAnon posts directly was calculated by retrieving the number of followers of a cluster that created each post and summing the numbers of followers across all posts in a given month. The sum can be viewed as the approximated total number of users who received QAnon posts on their social media timeline in a given month. (Users who were subscribing to two or more QAnon pages or groups need to be considered for a more accurate estimation, but information about these users is not available to researchers.) The result showed that the number of users directly exposed to QAnon posts was 85 million in November 2017, 365 million in August 2020, and 76 million in November 2020. The total number of users receiving QAnon posts in the entire time period was 3.1 billion.

*A.3   The number of posts in a year.*   For each year, we counted the number of posts generated by each active cluster (Fig. 1B). On average, an active cluster generated 4.03 posts in 2017 ($SD = 8.46$, $N = 524$), 14.05 posts in 2018 ($SD = 30.57$, $N = 1,464$), 13.10 posts in 2019 ($SD = 61.49$, $N = 1,804$), and 26.25 posts in 2020 ($SD = 50.30$, $N = 2,714$).

*A.4   Weekly posts by cohort.*   A cluster cohort was defined as a group of QAnon clusters that created their first post in the same year. All clusters were classified into four different cohorts: 2017, 2018, 2019, and 2020 cohorts. The average weekly number of posts was calculated by diving the number of posts from a cluster by the number of hours between its first and last post, and multiplying the result by 168 hours (24 hours × 7 days). This metric was calculated only for clusters that generated at least two QAnon posts. The average weekly number of posts was 0.45 for the 2017 cohort ($SD = 0.58$, $N = 524$), 0.35 posts for the 2018 cohort ($SD = 0.95$, $N = 965$), 0.85 posts for the 2019 cohort ($SD = 2.30$, $N = 549$), and 7.13 posts for the 2020 cohort ($SD = 151.63$, $N = 767$). See Fig. 1D.

*A.5   Early activity by cohort.*   The average number posts created by a cluster within the first month after the creation of its first QAnon post was 3.23 in the 2017 cohort ($SD = 6.15$, $N = 524$), 2.48 in the 2018 cohort ($SD = 6.24$, $N = 968$), 4.15 in the 2019 cohort ($SD = 19.18$, $N = 552$), and 8.81 in the 2020 cohort ($SD = 20.88$, $N = 769$). The results are shown in Fig. 1E.

## B    Appendix B: Information Sources for the QAnon Narrative

*B.1   Descriptive statistics.*   In total, 91.4% of the posts included at least one URL, while only 8.6% did not include any URL. When aggregated by clusters, 90.9% ($SD = 11.6\%$, $N = 2,813$) of posts in a QAnon cluster included one or more URLs. 75.8% of posts in a cluster ($SD = 18.6\%$, $N = 2,813$) used at least one of the four types of information sources: Facebook internal sources, social media sources, news sources, and low credibility sources.

An average cluster had 47.8% ($SD = 25.5\%$, $N = 2,813$) of posts using Facebook internal sources. The average proportions of posts in a cluster using non-Facebook social media sources, news sources, and low credibility sources were 25.8% ($SD = 21.7\%$, $N = 2,813$), 8.1% ($SD = 12.1\%$, $N = 2,813$), and 2.1% ($SD = 6.0\%$, $N = 2,813$), respectively.

The 20 most frequently used domains of the URLs were the following: facebook.com (31.2%), youtube.com (20.6%), twitter.com (6.3%), google.com (1.4%), amazon.com (1.3%), patreon.com (1.1%), gab.ai (0.7%), nativate3d.no (0.7%), paypal.me (0.6%), blogspot.com (0.6%), themillennialbridge.com (0.5%), foxnews.com (0.4%), bitchute.com (0.4%), wikipedia.org (0.4%), google.no (0.4%), wordpress.com (0.3%), nbcnews.com (0.3%), instagram.com (0.3%), stillnessinthestorm.com (0.3%), and conspiracydailyupdate.com (0.3%).

*B.2   Facebook internal sources.*   The proportion of posts using Facebook internal sources significantly increased every year between 2017 and 2020. As shown in Fig. 3A, the proportion of posts using Facebook internal sources in a cluster on average was 30.0% ($SD = 40.1\%$, $N = 524$) in 2017, 37.2% ($SD = 34.6\%$, $N = 1,464$) in 2018, 44.7% ($SD = 36.3\%$, $N = 1,804$) in 2019, and 50.0% ($SD = 27.5\%$, $N = 2,714$). Differences in median between 2017 and 2018 (Mann-Whitney $U = 310445.0$, $P < .001$), between 2018 and 2019 ($U = 1169646.5$, $P < .001$), and between 2019 and 2020 ($U = 2176166.5$, $P < .001$) were statistically significant. The mean differences between 2017 and 2018 (Welch's $t(1986) = 3.67$, $P < .001$), between 2018 and 2019 (Welch's $t(3266) = 5.96$, $P < .001$), and between 2019 and 2020 (Welch's $t(4516) = 5.33$, $P < .001$) were also statistically significant.

*B.3   External sources.*   Posts using external sources significantly decreased every year from 2017 to 2020 (Fig. 3A). The proportion of posts using these sources in a cluster on average was 74.4% ($SD = 37.7\%$, $N = 524$) in 2017, 65.8% ($SD = 33.5\%$, $N = 1,464$) in 2018, 59.1% ($SD = 35.2\%$, $N = 1,804$) in 2019, and 52.7% ($SD = 27.7\%$, $N = 2,714$) in 2020. Differences in median between 2017 and 2018 (Mann-Whitney $U = 295141.5$, $P < .001$), between 2018 and 2019 ($U = 1181651.0$, $P < .001$), and between 2019 and 2020 ($U = 2102013.5$, $P < .001$) were statistically

significant. The mean differences between 2017 and 2018 (Welch's $t(1986)$ = -4.61, $P < .001$), between 2018 and 2019 (Welch's $t(3266)$ = -5.50, $P < .001$), and between 2019 and 2020 (Welch's $t(4516)$ = -6.50, $P < .001$) were all statistically significant.

*B.4 Social media sources.* Posts using social media sources other than Facebook decreased every year significantly from 2017 to 2020 (Fig 3B). The proportion of posts using non-Facebook social media sources in a cluster on average was 40.6% ($SD$ = 42.3%, $N$ =524) in 2017, 32.1% ($SD$ = 32.9%, $N$ = 1,464) in 2018, 26.7% ($SD$ = 31.4%, $N$ = 1,804) in 2019, and 24.0% ($SD$ = 23.1%, $N$ = 2,714). Differences in median between 2017 and 2018 (Mann-Whitney $U$ = 359655.0, $P$ = .015), between 2018 and 2019 ($U$ = 1190729.5, $P < .001$), and between 2019 and 2020 ($U$ = 2334970.0, $P$ = .004) were statistically significant. The mean differences between 2017 and 2018 (Welch's $t(1986)$ = -4.17, $P < .001$), between 2018 and 2019 (Welch's $t(3266)$ = -4.82, $P < .001$), and between 2019 and 2020 (Welch's $t(4516)$ = -3.11, $P$ = .002) were all statistically significant.

*B.5 News sources.* The proportion of posts including news sources was 6.5% ($SD$ = 20.8%, $N$ = 524) in 2017, 8.3% ($SD$ = 19.0%, $N$ = 1,464) in 2018, 9.0% ($SD$ = 19.8%, $N$ = 1,804) in 2019, and 8.3% ($SD$ = 14.1%, $N$ = 2,714) in 2020, as shown in Fig. 3B. Differences between 2017 and 2018 (Mann-Whitney $U$ = 320218.0, $P < .001$) and between 2019 and 2020 ($U$ = 2142255.5, $P < .001$) were statistically significant, while the difference between 2018 and 2019 was not significant ($U$ = 1301024.5, $P$ = .193). The mean differences were not statistically significant: between 2017 and 2018 (Welch's $t(1986)$ = 1.78, $P$ = .076), between 2018 and 2019 (Welch's $t(3266)$ = 0.98, $P$ = .328), and between 2019 and 2020 (Welch's $t(4516)$ = -1.30, $P$ = .192).

*B.6 Low credibility sources.* QAnon clusters' reliance on low credibility sources was 5.0% ($SD$ = 18.2%, $N$ = 524) in 2017, 3.8% ($SD$ = 12.2%, $N$ = 1,464) in 2018, 3.4% ($SD$ = 12.3%, $N$ = 1,804) in 2019, and 1.3% ($SD$ = 6.1%, $N$ = 2,714) in 2020, as shown in Fig. 3B. Differences between 2017 and 2018 (Mann-Whitney $U$ = 346044.0, $P < .001$), between 2018 and 2019 ($U$ = 1273279.5, $P$ = .006), and between 2019 and 2020 ($U$ = 2356655.5, $P < .001$) were statistically significant. The mean difference between 2019 and 2020 (Welch's $t(4516)$ = -6.48, $P < .001$) was statistically significant, while the differences between 2017 and 2018 (Welch's $t(1986)$ = -1.42, $P$ = .156) and between 2018 and 2019 (Welch's $t(3266)$ = -0.99, $P$ = .324) were not significant.

*B.7 Comparison between Facebook videos and YouTube videos.* The two most frequently used sources for QAnon clusters were Facebook and YouTube, as reported in Appendix C.1, and these two sources accounted for 51.9% of all URLs. We compared these two sources as an important case analysis showing the difference between internal information sources within Facebook and the most dominant external source, YouTube.
For this purpose, we examined the proportion of posts based on Facebook videos and YouTube videos. Facebook categorizes each post into one of the following types: photo, link, native_video, youtube, video, status, live_video_complete, live_video, live_video_scheduled, and album. We called posts in the native_video, video, live_video_complete, live_video, or live_video_scheduled categories as "posts using Facebook videos." On the other hand, "posts using YouTube videos" referred to posts in the "youtube" category.

As depicted in Fig. 3E, the proportion of posts using YouTube videos in a cluster decreased from 2017 to 2020: 21.8% in 2017 ($SD = 35.1\%$, $N = 524$), 16.7% in 2018 ($SD = 26.3\%$, $N = 1,464$), 11.0% in 2019 ($SD = 22.3\%$, $N = 1,804$), and 10.1% in 2020 ($SD = 15.4\%$, $N = 2,714$). Differences between 2017 and 2018 (Mann-Whitney $U = 338553.5$, $P < .001$) and between 2019 and 2020 ($U = 1943618.0$, $P < .001$) were statistically significant, while the difference between 2018 and 2019 was not significant ($U = 1309299.5$, $P = .331$).

Contrarily, the proportion of posts using Facebook videos in a cluster declined. It was 21.8% in 2017 ($SD = 35.2\%$, $N = 524$), 22.0% in 2018 ($SD = 29.2\%$, $N = 1,464$), 22.6% in 2019 ($SD = 29.2\%$, $N = 1,804$), and 27.0% in 2020 ($SD = 22.9\%$, $N = 2,714$). Differences between 2018 and 2019 (Mann-Whitney $U = 1146920.0$, $P < .001$) and between 2019 and 2020 ($U = 2133890.5$, $P < .001$) were statistically significant, while the difference between 2017 and 2018 was not ($U = 369576.5$, $P = .087$).

## C    Appendix C: Association between Information Sources and User Engagement

To estimate the association between information source types on user engagement, we fit negative binomial regression models with maximum likelihood estimation (MLE) to consider overdispersion [6, 25]. For one of the most prominent examples, Brady et al. [6] used negative binomial regression models with MLE to analyze how user engagement on Twitter was influenced by the inclusion of moral-emotional language in social media messages. We also fit Poisson models with MLE to check the robustness of the findings. The dependent variables were the number of shares, the number of comments, and the number of reactions to a post. The number of reactions was defined as the sum of 8 different reaction counters of a post: like, love, wow, haha, sad, angry, thankful, and care.

In this section, we present the specifications and results of two analyses. First, among posts with at least one information source, we examined the association between user engagement, and the inclusion of Facebook internal sources and the inclusion of external sources in a post, respectively. Second, among posts with at least one information source, we estimated the association between the subtypes of information sources (Facebook internal, social media, news, and low credibility sources) and user engagement and additional results for a robustness check. All standard errors were clustered at the QAnon cluster level, and all models were estimated with cluster-robust standard error at the QAnon cluster level. Incidence rate ratios (*IRR*s) were calculated by exponentiating the coefficients of each regression model. SciPy (version 1.6.1), a package based on the Python programming language, was used for all analyses. The data and the script for regression analyses are available online [49].

*C.1    Association between internal/external sources and user engagement.* Among posts with at least one information source ($N = 107,392$), we examined the associations between the existence of Facebook internal sources and user engagement and between the existence of external sources and user engagement. The first model predicting the number of shares (52% of posts were shared more than once) included two binary predictors indicating the inclusion of Facebook internal sources in a post and the inclusion of external sources, and five covariates: (1) cluster size (a counting variable representing the number of Facebook users following the cluster), (2) posting month (a continuous variable ranging between 0 and 106), (3) the existence of videos in a post (a binary variable indicating whether or not a post includes at least one video), (4) the existence of photos in a post (a binary variable indicating whether or not a post includes at least one photo), (5) cluster type (a categorical variable indicating the type of a given cluster: a page or a group). The second model predicting the number of comments in a post (36% of posts had more than one comment) and the third model predicting the number of reactions in a post (70% of posts had more than one reaction) had the same two predictors and five covariates as the first model.

The regression results showed that QAnon posts using Facebook internal sources attracted more shares ($IRR = 1.68$, 95% CI [1.20, 2.35], $P = .002$) and comments ($IRR = 1.86$, 95% CI [1.09, 3.19], $P = .024$). Statistical significance was not identified for the association between the internal sources and reactions ($IRR = 1.24$, 95% CI [0.87, 1.75], $P = .231$). QAnon posts using external sources, on the other hand, attracted fewer shares ($IRR = 0.61$, 95% CI [0.46, 0.82], $P = .001$), comments ($IRR = 0.59$, 95% CI [0.38, 0.91], $P = .017$), and reactions ($IRR = 0.50$, 95% CI [0.38, 0.65], $P < .001$) than those not using external sources.

*C.2    Association between source subtypes and user engagement.* Among posts with at least one information source ($N = 107,392$), we examined the association between information source types and user engagement. The first model predicting the number of shares included four binary predictors indicating the inclusion of different information source types in a post (Facebook internal source, social media source, news source, and low credibility content source) and the same five covariates explained in Appendix D.1. The second model predicting the number of comments in a post and the third model predicting the number of reactions in a post had the same predictors and covariates as the first model.

The regression results are reported in Table 3 and visualized in Fig. 3.

Table 3. The Association Between the Use of Information Sources and User Engagement

| Independent variables | Dependent variable | | |
|---|---|---|---|
| | Share | Comment | Reaction |
| FB internal | 0.482 ** | 0.611 ** | 0.280 |
| News | 0.465 * | 0.509 *** | 0.400 *** |
| Social media | -0.653 *** | -0.695 *** | -0.772 *** |
| Low credibility | 0.271 | -0.112 | -0.049 |
| Month | 0.038 *** | 0.033 *** | 0.037 *** |
| FB group (ref: page) | -1.498 *** | -0.676 ** | -1.118 *** |
| Subscriber | 5.61e-6 *** | 7.14e-6 *** | 8.98e-6 *** |
| Photo | -0.006 | -0.112 | 0.378 |
| Video | -0.068 | -0.058 | -0.048 |

Note: Results of negative binomial models estimating the number of shares, comments, and reactions of a post as a function of variables shown in the leftmost

*C.3   Robustness check.*   To check the robustness of the findings, the association between
information source types and user engagement was examined using Poisson regression models.
Dependent and independent variables of the Poisson models were identical to those of the
Negative Binomial models mentioned in D.2. Results from these Poisson models could be
informative in checking the robustness of the statistical findings presented in D.2, but these
models could not be an optimal choice for the present data due to the overdispersion of the
dependent variables.

This robustness check confirmed that QAnon posts using Facebook internal sources attracted
more shares (*IRR* = 2.84, 95% CI [1.74, 4.63], *P* < .001) and comments (*IRR* = 1.83, 95% CI
[1.16, 2.90], *P* = .010) than those not using internal sources. The association between the use of
Facebook internal sources and reactions was not statistically significant (*IRR* = 1.50, 95% CI
[0.94, 2.40], *P* = .091).

The inclusion of social media sources was negatively associated with shares (*IRR* = 0.38, 95%
CI [0.26, 0.56], *P* < .001), comments (*IRR* = 0.32, 95% CI [0.18, 0.56], *P* < .001), and reactions (*IRR*
= 0.23, 95% CI [0.12, 0.45], *P* < .001).

The association between the use of news sources and shares (*IRR* = 1.44, 95% CI [0.99, 2.09], *P* =
.058), comments (*IRR* = 1.27, 95% CI [0.92, 1.76], *P* = .150), and reactions (*IRR* = 0.96, 95% CI
[0.59, 1.57], *P* = .872) were not statistically significant.

In this analysis for a robustness check, the association of including low credibility sources in a
post with the number of shares (*IRR* = 0.23, 95% CI [0.08, 0.66], *P* = .007), the number of
comments (*IRR* = 0.28, 95% CI [0.12, 0.64], *P* = .003), and the number of reactions (*IRR* = 0.23,
95% CI [0.09, 0.64], *P* = .004) appeared to be negative significantly.